\documentclass[a4paper,11pt]{article}
\usepackage{pos}
\usepackage{wrapfig}
\usepackage{enumitem}
\newcommand{\doiref}[2]{\href{https://doi.org/#1}{#2}}

\patchcmd{\thebibliography}{\settowidth}{\setlength{\itemsep}{0pt plus 0.1ex}\settowidth}{}{}

\title{%
    Latest results from the searches for ultra-high-energy
    photons at the Pierre Auger Observatory}
 \ShortTitle{Latest results from the searches for UHE-photons at the Pierre Auger Observatory}

\author*[ab]{Pierpaolo Savina}

\affiliation[a]{%
Gran Sasso Science Institute (GSSI), Via Iacobucci 2, I-67100 L’Aquila, Italy}
\affiliation[b]{
Istituto Nazionale di Fisica Nucleare (INFN) - Laboratori Nazionali del Gran Sasso, Via G. Acitelli 22, I-67100 Assergi, L’Aquila, Italy}

\onbehalf{for the Pierre Auger Collaboration$^c$}
\affiliation[c]{Observatorio Pierre Auger, Av.\ San Mart{\'\i}n Norte 304, 5613 Malarg\"ue, Argentina\\
Full author list: {\rm\url{https://www.auger.org/archive/authors_icrc_2025.html}}}

\emailAdd{spokespersons@auger.org}

\abstract{%
The Pierre Auger Observatory is the largest air-shower detector in the world, offering unparalleled exposure to photons with energies above $5\times10^{16}$\,eV.
Since the start of data collection almost two decades ago, numerous searches for photons have been conducted using the detection systems of the Observatory.
These searches have led to the most stringent upper limits on the diffuse photon flux.
These limits place severe constraints on current models regarding the origin of ultra-high-energy cosmic rays and emphasize the significant capabilities of the Pierre Auger Observatory in the context of multimessenger astronomy at the highest energies.
This contribution provides an overview of the ongoing efforts to search for high-energy photons in the data from the Pierre Auger Observatory.
The latest results from searches for the diffuse photon flux will be presented, along with follow-up investigations for photons associated with transient events, such as gravitational wave detections.
Furthermore, future prospects will be discussed in light of the ongoing AugerPrime detector upgrade, which will enhance the sensitivity of the Observatory to photons up to the highest energies.
}

\ConferenceLogo{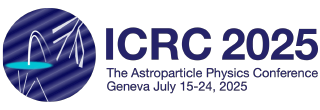}

\FullConference{%
39th International Cosmic Ray Conference (ICRC2025)\\
15 -- 24 July, 2025\\
Geneva, Switzerland}

\begin{document}
\maketitle

\section{Introduction}

The search for ultra-high-energy (UHE) photons of cosmic origin is one of the primary scientific goals of the Pierre Auger Observatory. 
While no UHE photons have been unambiguously identified to date, the upper limits on their flux have already placed strong constraints. 
The detection of UHE photons would be of fundamental importance for advancing the multimessenger approach to understanding the most extreme astrophysical phenomena. Unlike charged particles, neutral messengers such as photons point back directly to their sources, offering valuable directional information. However, unlike neutrinos, UHE photons interact with the background photon fields that permeate the Universe. These interactions significantly reduce their attenuation length to about 30~kpc at $10^{15}$~eV amd approximately 10~Mpc at $10^{19}$~eV.

In this contribution, we review the current status of UHE photon searches at the Pierre Auger Observatory. After a brief overview of the characteristics of photon-induced air showers (Sec.~\ref{sec::signatures}), we describe the Pierre Auger Observatory and its detection capabilities (Sec.~\ref{sec::pao}). We then focus on the search for a diffuse flux of UHE photons using the Observatory's various detector systems (Sec.~\ref{sec::diffuse}), followed by searches for UHE photons originating from transient events (Sec.~\ref{sec::followup}). 

\section{Photon-induced air Showers}
\label{sec::signatures}

There are two principal physical differences between air showers initiated by photons and those initiated by hadrons~\cite{photonreview2007}: first, the longitudinal development of a photon-induced air shower is delayed compared to that of hadron-induced showers. This delay arises from the lower multiplicity of electromagnetic interactions—which dominate in photon showers—relative to hadronic interactions. As a result, the atmospheric depth at which the number of shower particles reaches its maximum, $X_{\mathrm{max}}$, is typically greater for photon-induced showers. 
Moreover, due to the much larger mean free path for photo-nuclear interactions compared to the radiation length, only a small fraction of the electromagnetic component in a photon-induced shower is converted into the hadronic component, and subsequently into muons. As a result, these showers contain significantly fewer muons than hadron-induced showers of the same primary energy.
All searches for UHE photons using air-shower data are based on these two distinguishing features. $X_{\mathrm{max}}$ can be measured directly via the air-fluorescence technique. Although the water Cherenkov detectors of the Pierre Auger Observatory cannot yet directly measure the number of muons, it is possible to infer information about the muonic content by studying the lateral distribution of secondary particles at ground level. The steepness of this lateral distribution depends on both the muon number and the shower development profile, and is sensitive to the type of primary particle.

\section{The Pierre Auger Observatory}
\label{sec::pao}

The Pierre Auger Observatory~\cite{AugerNIM} is the largest cosmic-ray observatory currently in operation, offering unprecedented exposure to UHE photons. A defining feature of the Observatory is its hybrid detection strategy, which combines a Surface Detector array (SD) with a Fluorescence Detector (FD).
The SD consists of 1600 water-Cherenkov detectors arranged on a triangular grid with 1500~m spacing (referred to as SD-1500), covering an area of approximately 3000~km$^2$. Surrounding the array at four peripheral sites are 24 fluorescence telescopes that constitute the FD. While the SD continuously samples the lateral distribution of shower particles at ground level with an almost 100\% duty cycle, the FD records the longitudinal development of extensive air showers in the atmosphere. However, the FD operates only during clear, moonless nights, limiting its duty cycle to about 15\%.
By combining data from both systems in so-called hybrid events, the Pierre Auger Observatory achieves superior accuracy in reconstructing air-shower parameters compared to using either system alone.
To extend the sensitivity of the Observatory to lower energies (below $10^{18}$\,eV), two denser infill sub-arrays have been deployed in the western section of the SD: \textbf{SD-750}, composed of 50 additional detectors placed between the standard SD-1500 stations with 750~m spacing, covering an area of approximately 27.5~km$^2$, and \textbf{SD-433}, a smaller sub-array consisting of 10 detectors with 433~m spacing, primarily used for calibration and dedicated low-energy studies.
To enable hybrid measurements in this lower-energy regime, where air showers typically develop higher in the atmosphere beyond the field of view of the standard FD telescopes, the SD-750 sub-array is operated in conjunction with the High-Elevation Auger Telescopes (HEAT) installed at the Coihueco FD site. The HEAT telescopes observe the atmosphere at elevation angles from $30^\circ$ to $60^\circ$, complementing the original Coihueco telescopes, which cover the range from $0^\circ$ to $30^\circ$.

\section{Searches for a diffuse flux of photons}
\label{sec::diffuse}

In this section, we summarize the four most recent diffuse,
i.e. direction-independent, unresolved, photon searches conducted at the Pierre Auger Observatory, ordered by increasing energy range. At $E > 5 \times 10^{16}$\,eV, a dedicated analysis using underground muon detectors (UMD) and the SD-433 has been performed~\cite{AugerSD-433}. In the range $2 \times 10^{17}$\,eV $< E < 10^{18}$\,eV, a low-energy hybrid analysis combining $X_{\max}$ and SD observables was carried out~\cite{AugerHECO}. At  $E > 10^{18}$\,eV, an updated hybrid analysis was used~\cite{AugerFD}, while for $E > 10^{19}$\,eV, photon-induced air showers were searched for using SD data only~\cite{AugerSD}.

\subsection{\texorpdfstring{Energy range $\mathbf{E > 5\times10^{16}}$\,eV}{The E above 20 PeV energy range}}

\begin{figure}[ht]
  \centering
  \begin{minipage}{0.55\textwidth}
    \centering
    \includegraphics[width=\linewidth]{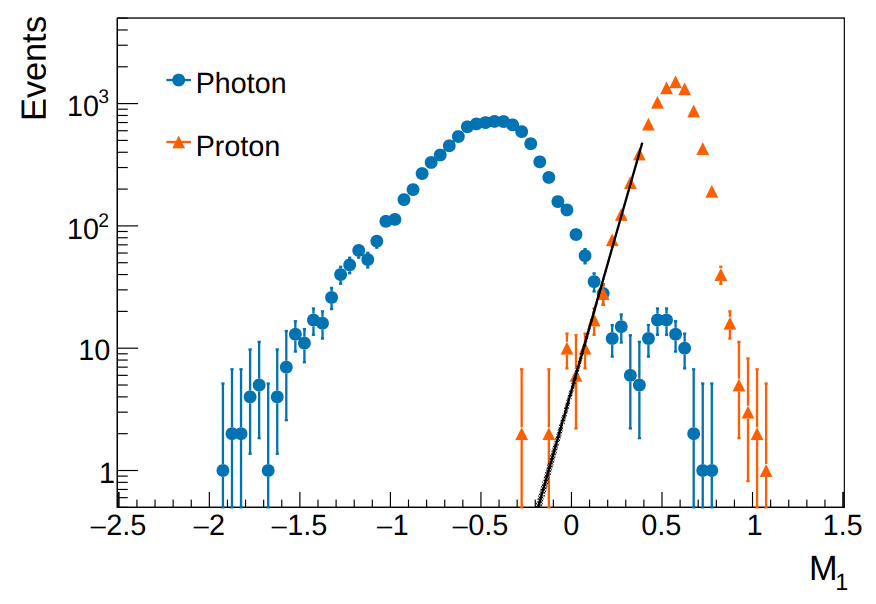}
  \end{minipage}%
  \hfill
  \begin{minipage}{0.4\textwidth}
    \caption{
      Distributions of $M_1$ for simulated photon- and proton-induced events in the energy range $\log_{10}(E_{\gamma,\mathrm{eq}}/\mathrm{eV}) \in (16.7, 16.9)$. Uncertainties in bins with fewer than ten entries correspond to 95\% confidence intervals using the Feldman--Cousins method. The black band shows the fit to the proton tail used to estimate background contamination. See details in ~\cite{AugerSD-433}.
    }
    \label{fig:M1Distr}
  \end{minipage}
    \vspace{-10pt}
\end{figure}

The photon search above $5 \times 10^{16}$\,eV exploits the low muon content of electromagnetic air showers, using signals from the UMD. A muon-based observable, 
$M_b = \log_{10} \left( \sum_i \frac{\rho_i}{\rho_{\mathrm{pr}}} \left( \frac{r_i}{r_{\mathrm{pr}}} \right)^b \right)$,
is constructed from the muon densities $\rho_i$ at stations $i$, normalized to the average proton-induced muon density $\rho_{\mathrm{pr}}$ at a reference distance $r_{\mathrm{pr}} = 200$\,m. This normalization absorbs energy and zenith-angle dependencies, where $\rho_{\mathrm{pr}}$ is
obtained from simulations.
To suppress fluctuations from peripheral stations, $M_b$ is computed using only the six detectors around the station with the highest signal. 
The parameter $b$ is optimized for photon-hadron separation, yielding $b=1$ and defining the final observable $M_1$. Photon-induced showers typically yield lower $M_1$ values, while muon-rich proton showers dominate the background.

As shown in Fig.~\ref{fig:M1Distr}, the $M_1$ distributions show a clear but non-perfect separation: rare muon-poor proton events and photonuclear interactions may cause overlap. A conservative estimate of background contamination is derived by fitting the lower tail of the proton $M_1$ distribution, while signal efficiency is defined relative to photon simulations. This approach enables a tunable trade-off between efficiency and contamination, crucial for rare photon searches.
Upper limits on the integral photon flux $\Phi^{\mathrm{UL}}_\gamma(E_\gamma > E_0)$ at 95\% confidence level are derived for threshold energies of 50, 80, 120, and 200\,PeV. The resulting upper limits on the integral flux are 12.3, 11.7, 11.3, and 11.3\,km$^{-2}$\,sr$^{-1}$\,yr$^{-1}$, respectively~\cite{AugerSD-433}.

\subsection{\texorpdfstring{%
Energy range $\mathbf{2 \times 10^{17}}$\,eV $\mathbf{< E < 10^{18}}$\,eV%
}{The 0.2 EeV to 1EeV range}}

\begin{wrapfigure}{r}{0.55\textwidth}
  \centering
  \vspace{-10pt}
  \includegraphics[width=0.53\textwidth]{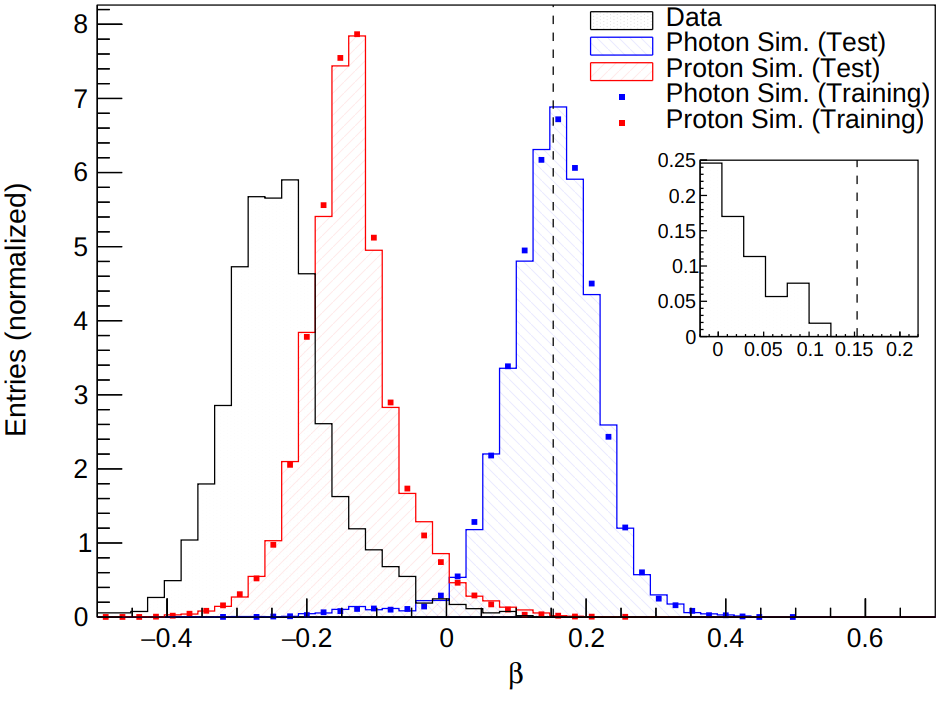}
  \vspace{-10pt}
  \caption{Distribution of simulated photons (blue), simulated protons (red), and data (black) of the BDT discriminator, $\beta$, combining $X_{\max}$, $S_b$, number of stations, photon energy, and zenith. The selection threshold for photon candidates is shown by the dashed line.}
  \label{fig:BDTDistr}
\end{wrapfigure}

The photon search in the energy range, starting at $2 \times 10^{17}$\,eV~\cite{AugerHECO}, uses data from the SD-750 array combined with HEAT and Coihueco telescopes (HeCo). Three observables are used in a multivariate analysis based on Boosted Decision Trees (BDT): $X_{\text{max}}$ (from FD), the signal-based parameter $S_b = \sum_i S_i (r_i / 1000~\text{m})^b$ with $b=4$ (sensitive to shower development and muon content), and $N_{\text{stations}}$, the number of triggered SD detectors. Simulated samples of primary photons (signal) and protons (background) are used for training and performance evaluation. The BDT includes also $E_\gamma$ (calorimetric energy with 1\% correction for missing energy) and the reconstructed zenith angle $\theta$.
The analysis was applied to hybrid HeCo+SD-750 data collected from June 2010 to December 2015, yielding 2204 selected events above $2 \times 10^{17}$\,eV. No photon candidates are found above the BDT cut (50\% signal efficiency), with a background rejection of $(99.91 \pm 0.03)\%$.
Upper limits on the integral photon flux at 95\% CL are derived assuming a $E^{-2}$ spectrum and  an  efficiency-weighted exposure of 2.4–2.7\,km$^2$\,sr\,yr. The limits are 2.72, 2.50, 2.74, and 3.55\,km$^{-2}$\,sr$^{-1}$\,yr$^{-1}$ for threshold energies of $2$, $3$, $5 \times 10^{17}$\,eV, and $10^{18}$\,eV, respectively. These translate to upper limits on the photon fraction of 0.28\%, 0.63\%, 2.20\%, and 13.8\%~\cite{AugerHECO}.

\subsection{\texorpdfstring{Energy range $\mathbf{E > 10^{18}}$\,eV}{The E above 1 EeV energy range}}

The photon search at energies above $10^{18}$\,eV~\cite{AugerFD} is performed using hybrid events. As in the lower-energy analysis, $X_{\text{max}}$ is obtained from FD measurements, while the muon content of air showers is quantified by the parameter $F_\mu$, derived from SD signals using air-shower universality~\cite{AugerFD}. 

\begin{wrapfigure}{l}{0.55\textwidth}
  \centering
  \vspace{-10pt}
  \includegraphics[width=0.53\textwidth]{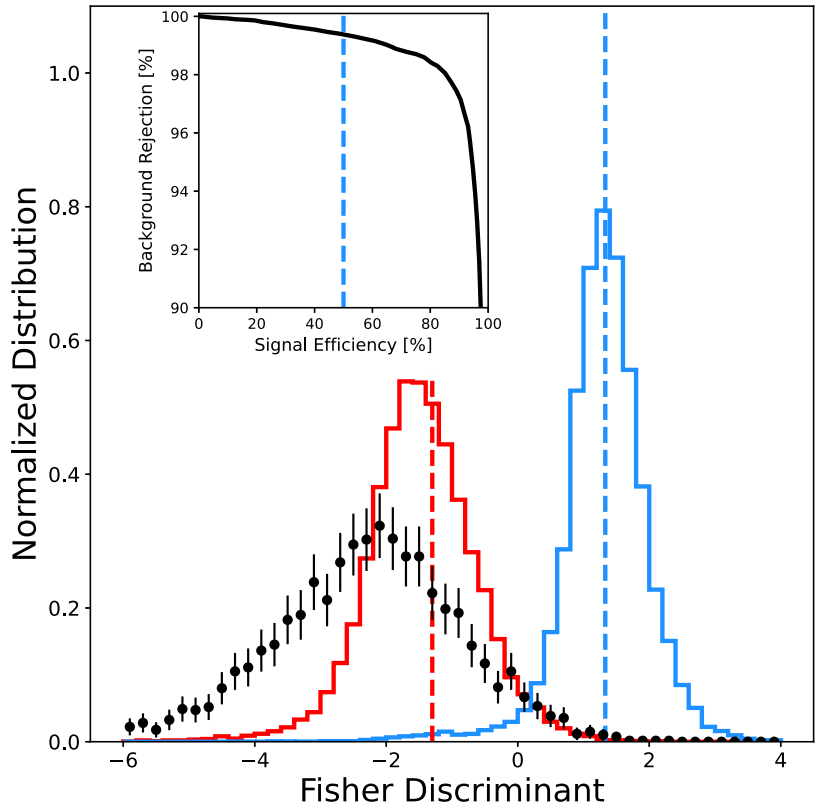}
  \vspace{-10pt}
\caption{Distribution of the Fisher discriminant for simulated photons (signal, blue), protons (background, red), and for the burnt sample (black). The vertical red line marks the threshold $f = -1.3$, above which the background decreases nearly exponentially while the photon selection efficiency remains close to 100\%. The blue line indicates the median of the photon distribution. Inset: background rejection as a function of signal efficiency from the Fisher discriminant analysis.}
  \label{fig:FisherUniversality}
\end{wrapfigure}

$F_\mu$ is a proxy of the muon content of the shower. it is derived 
for each hybrid event by matching
the observed SD signals to the signal predicted by the universality model.
Combined with $X_{\text{max}}$ and the photon-estimated energy $E_\gamma$, $F_\mu$ is used in a linear Fisher discriminant analysis to separate photons from hadrons (Fig.~\ref{fig:FisherUniversality}).
The Fisher discriminant $f$ shows strong separation between simulated photon and proton events. A 5\% burnt sample of data is used to estimate the background. Due to limited statistics, only events with $f > -1.3$ are used to model the tail of the background, which is then fitted and scaled to the full data set (2005–2017, $\sim$32,000 hybrid events above $10^{18}$\,eV). After applying the photon selection cut (at the median of the photon $f$ distribution), 22 photon candidates are identified, consistent with the background expectation of $30 \pm 15$ events.
Upper limits on the integral photon flux $\Phi^{\text{UL}}_\gamma(E_\gamma > E_0)$ at 95\% CL are set for thresholds of $1$, $2$, $3$, $5 \times 10^{18}$\,eV, and $10^{19}$\,eV. The number of photon candidates observed for each threshold is 22, 2, 0, 0, and 0, respectively. These are compatible with the background expectations $30 \pm 15$, $6 \pm 6$, $0.7 \pm 1.9$, $0.06 \pm 0.25$, and $0.02 \pm 0.06$. The efficiency-weighted exposure, derived from simulations assuming a $E^{-2}$ spectrum, ranges from 420.7 to 1245.9\,km$^2$\,sr\,yr. The resulting upper limits on the integral photon flux are: 4.0, 1.1, 0.35, 0.23, and 0.0021\,km$^{-2}$\,sr$^{-1}$\,yr$^{-1}$~\cite{AugerFD}.

\subsection{\texorpdfstring{%
Energy range $\mathbf{E > 10^{19}}$\,eV%
}{The above 10 EeV energy range}}

\begin{figure}[ht]
  \centering
  \begin{minipage}{0.55\textwidth}
    \centering
    \includegraphics[width=\linewidth]{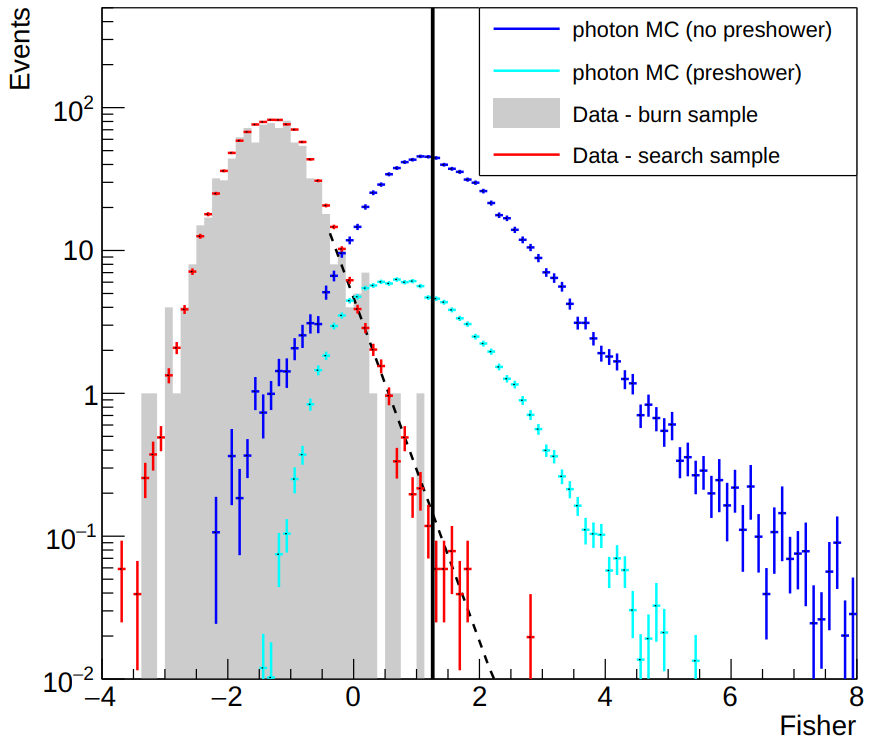}
  \end{minipage}%
  \hfill
  \begin{minipage}{0.4\textwidth}
    \caption{
      Distributions of the Fisher discriminant for the burnt sample (grey), the search sample (red), and simulated primary photons—non-preshowering (blue) and preshowering (light blue)—weighted with an $E^{-2}$ spectrum. The search and photon distributions are rescaled to match the integral of the burnt sample. The vertical line indicates the photon-candidate cut. The dashed line represents an exponential fit to the top 5\% of events in the burnt sample with the highest Fisher discriminant values. See details in~\cite{AugerSD}.
    }
    \label{fig:FisherSD}
  \end{minipage}
  \vspace{-20pt}
\end{figure}

At energies above $10^{19}$\,eV, the data collected by the SD-1500 are considered. While the high duty cycle of the SD provides a large exposure, the absence of FD measurements poses challenges, particularly for the determination of $X_{\text{max}}$.
Two observables are employed for photon-hadron discrimination: one related to the thickness of the shower front, and one based on the steepness of the lateral distribution. The first observable, $\Delta$, is derived from the risetime $t_{1/2}$ in individual SD stations—the time interval in which the integrated signal rises from 10\% to 50\% of its total value. Photon-induced showers typically exhibit longer risetimes due to their lower muon content and deeper development in the atmosphere. 
The second observable, $L_\text{LDF}$, quantifies the steepness of the lateral distribution function (LDF) of the signal. 
The two observables are combined into a Fisher discriminant. The analysis is applied to SD data collected between 1 January 2004 and 30 June 2020, including only events with zenith angles between 30$^\circ$ and 60$^\circ$ to ensure full shower development before ground. After applying quality cuts~\cite{AugerSD}, the search sample includes 48,061 events with $E_\gamma \geq 10^{19}$\,eV; a burnt sample of 886 events (1.8\%) is used to estimate the background.
Applying the photon-candidate cut (defined as the median of the Fisher distribution for simulated non-preshowering photons), 16, 2, and 0 events are observed above $1$, $2$, and $4 \times 10^{19}$\,eV, respectively. These results are consistent with expectations from an exponential fit to the tail of the burn-sample Fisher distribution. No excess or peak-like feature indicative of a photon signal is observed.
Upper limits on the integral photon flux are derived at 95\% CL, using simulation-based signal efficiencies that increase from 0.26 to 0.39 for thresholds between $10^{19}$\,eV and $4 \times 10^{19}$\,eV. The resulting upper limits on the photon flux are:
$2.11 \times 10^{-3}$\,km$^{-2}$\,sr$^{-1}$\,yr$^{-1}$ for $E_\gamma > 10^{19}$\,eV,
$0.312 \times 10^{-3}$\,km$^{-2}$\,sr$^{-1}$\,yr$^{-1}$ for $E_\gamma > 2 \times 10^{19}$\,eV, and
$0.172 \times 10^{-3}$\,km$^{-2}$\,sr$^{-1}$\,yr$^{-1}$ for $E_\gamma > 4 \times 10^{19}$\,eV.

\subsection{Results from the diffuse analyses}
\begin{figure}
    \centering
    \includegraphics[width=0.82\linewidth]{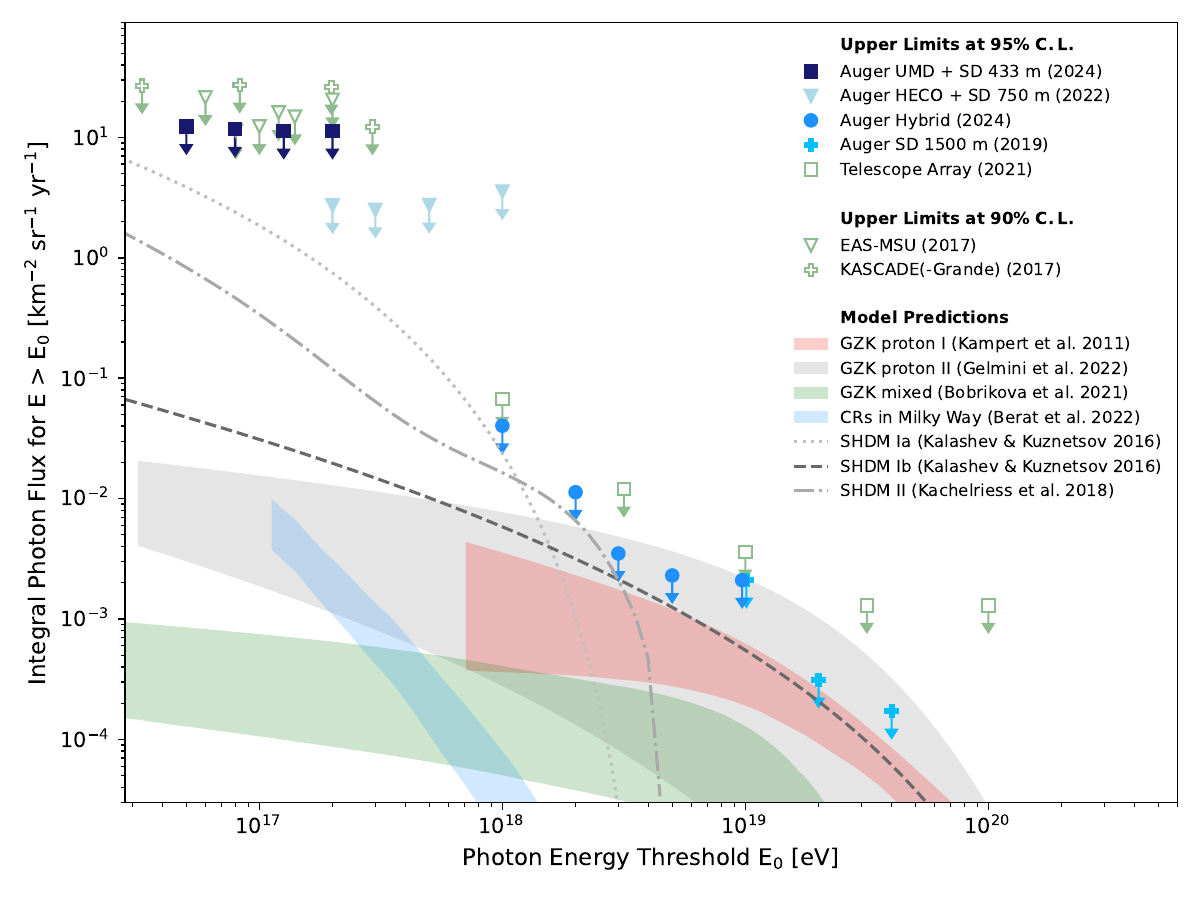}
\caption{Upper limits on the integral photon flux above threshold energy $E^{\mathrm{th}}_\gamma$ from this work (red markers, 95\% CL) and previous Pierre Auger results at higher energies (blue and black markers, 95\% CL), alongside limits from other experiments (90\% CL, except Telescope Array at 95\%). Shaded bands show cosmogenic flux predictions from UHECR interactions with galactic matter (gray), background radiation fields (violet, green, orange), and hot gas in the galactic halo (blue). Dashed lines denote super-heavy dark matter predictions (more details in~\cite{AugerSD-433}).}
    \label{fig:UL}
\end{figure}

The upper limits on the integral photon flux obtained from the four analyses (see Fig.~\ref{fig:UL}) represent the most stringent constraints to date across $2 \times 10^{16}$\,eV to the highest energies. The results are robust against systematic uncertainties. Fig.~\ref{fig:UL} also compares expectations from cosmogenic photons (proton and mixed-composition models), Galactic interactions, and super-heavy dark matter (SHDM) decay. Current limits start to constrain optimistic proton scenarios, while SHDM models remain viable only within specific mass–lifetime combinations.

\section{Follow-up studies to transient events}
\label{sec::followup}

\begin{figure}
    \centering
    \includegraphics[width=\linewidth]{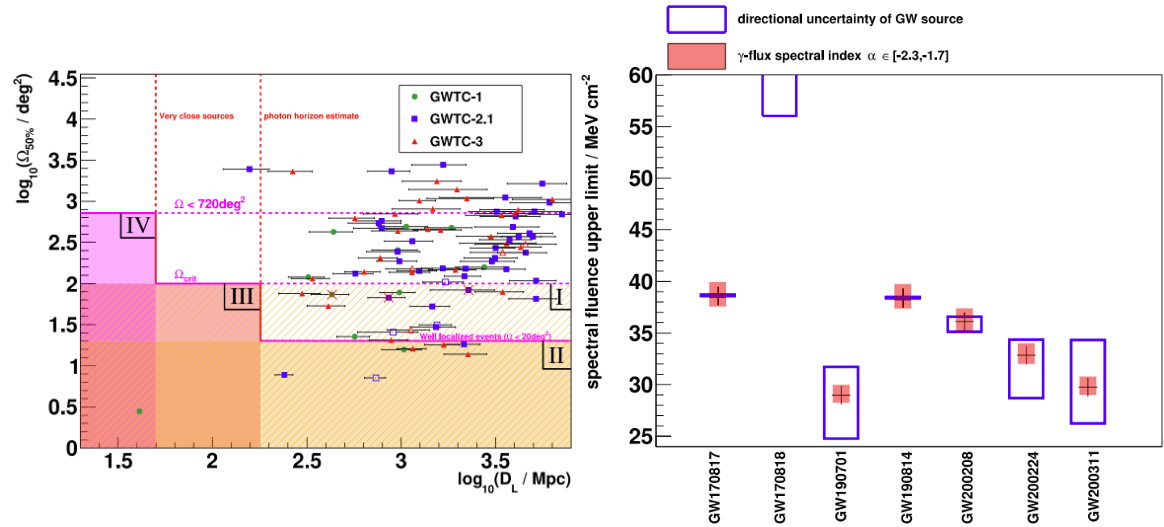}
    \caption{Left: Three classes of gravitational wave sources in the photon follow-up search, defined by their 50\% localization region ($\Omega_{50\%}$) and luminosity distance ($D_L$). Circled markers indicate events overlapping with the SD field of view. Right: Upper limits on the spectral fluence of UHE photons at Earth for each source; error bars include directional uncertainty (blue) and spectral index uncertainty (red). For the second event, error bars exceed the plot range due to its position near the SD field of view edge~\cite{GW}.}

    \label{fig:FollowUPGW}
\end{figure}

Detecting UHE photons from distant sources poses considerable challenges. The flux is expected to be strongly attenuated through interactions with cosmic background radiation fields, and air showers initiated by hadronic cosmic rays create a substantial background that complicates unambiguous photon identification. Photon candidate events are selected using standard criteria~\cite{AugerSD} applied to SD-1500 data.
To maximize sensitivity to transient photon sources while suppressing background, a dedicated selection strategy has been developed for gravitational wave follow-up. This strategy accepts only GW events that are either nearby or well-localized on the sky. Three classes of acceptable GW events are defined in the plane of 50\% sky localization region versus luminosity distance. Nearby sources are most likely to produce detectable UHE photons, but the detection of photons from distant, well-localized events could indicate physics beyond the Standard Model.
From the GWTC-1 and GWTC-2 catalogs, 10 gravitational wave events—including GW170817, associated with a neutron star merger and a coincident short gamma-ray burst were selected for targeted photon searches. The analysis covered a time window of one sidereal day following each GW event. No photon candidate events were detected. 
A similar search strategy was applied to investigate potential UHE photon emission associated with other transient sources, such as the blazar TXS 0506+056. No photon candidates were observed during either of the two periods of enhanced neutrino activity reported by IceCube: October 2014–February 2015 and March–September 2017.
At gigaparsec distances, the non-detection of UHE photons is consistent with expectations based on standard photon propagation models. The observation of such photons would likely require new physics, such as reduced photon attenuation in the extragalactic medium.

\section{Conclusions}
The Pierre Auger Observatory provides unmatched exposure to ultra-high energy (UHE) cosmic rays, photons. It has set the most stringent upper limits on diffuse UHE photon fluxes and performed targeted searches in coincidence with gravitational waves and other transients, reinforcing its role in multimessenger astronomy at the highest energies.
The ongoing upgrade, \textit{AugerPrime}, will enhance these capabilities. Scintillator detectors are being added to each surface station to improve shower component separation and composition sensitivity. The stations are also being equipped with radio antennas, and upgraded electronics allow for improved timing and dynamic range. These developments will strengthen the Observatory’s ability to detect or further constrain UHE photons.


%
%
%

\clearpage
\section*{The Pierre Auger Collaboration}

{\footnotesize\setlength{\baselineskip}{10pt}
\noindent
\begin{wrapfigure}[11]{l}{0.12\linewidth}
\vspace{-4pt}
\includegraphics[width=0.98\linewidth]{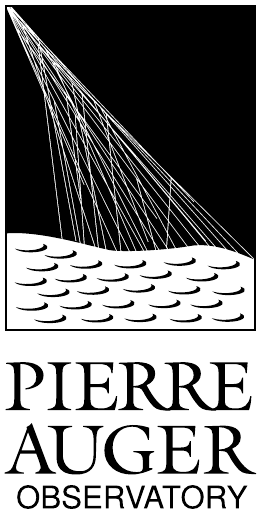}
\end{wrapfigure}
\begin{sloppypar}\noindent
A.~Abdul Halim$^{13}$,
P.~Abreu$^{70}$,
M.~Aglietta$^{53,51}$,
I.~Allekotte$^{1}$,
K.~Almeida Cheminant$^{78,77}$,
A.~Almela$^{7,12}$,
R.~Aloisio$^{44,45}$,
J.~Alvarez-Mu\~niz$^{76}$,
A.~Ambrosone$^{44}$,
J.~Ammerman Yebra$^{76}$
% ...

\end{sloppypar}
\begin{center}
\end{center}

\vspace{1ex}
\begin{description}[labelsep=0.2em,align=right,labelwidth=0.7em,labelindent=0em,leftmargin=2em,noitemsep,before={\renewcommand\makelabel[1]{##1 }}]
\item[$^{1}$] Centro At\'omico Bariloche and Instituto Balseiro (CNEA-UNCuyo-CONICET), San Carlos de Bariloche, Argentina
\item[$^{2}$] Departamento de F\'\i{}sica and Departamento de Ciencias de la Atm\'osfera y los Oc\'eanos, FCEyN, Universidad de Buenos Aires and CONICET, Buenos Aires, Argentina
\item[$^{3}$] IFLP, Universidad Nacional de La Plata and CONICET, La Plata, Argentina
\item[$^{4}$] Instituto de Astronom\'\i{}a y F\'\i{}sica del Espacio (IAFE, CONICET-UBA), Buenos Aires, Argentina
\item[$^{5}$] Instituto de F\'\i{}sica de Rosario (IFIR) -- CONICET/U.N.R.\ and Facultad de Ciencias Bioqu\'\i{}micas y Farmac\'euticas U.N.R., Rosario, Argentina
\item[$^{6}$] Instituto de Tecnolog\'\i{}as en Detecci\'on y Astropart\'\i{}culas (CNEA, CONICET, UNSAM), and Universidad Tecnol\'ogica Nacional -- Facultad Regional Mendoza (CONICET/CNEA), Mendoza, Argentina
\item[$^{7}$] Instituto de Tecnolog\'\i{}as en Detecci\'on y Astropart\'\i{}culas (CNEA, CONICET, UNSAM), Buenos Aires, Argentina
\item[$^{8}$] International Center of Advanced Studies and Instituto de Ciencias F\'\i{}sicas, ECyT-UNSAM and CONICET, Campus Miguelete -- San Mart\'\i{}n, Buenos Aires, Argentina
\item[$^{9}$] Laboratorio Atm\'osfera -- Departamento de Investigaciones en L\'aseres y sus Aplicaciones -- UNIDEF (CITEDEF-CONICET), Argentina
\item[$^{10}$] Observatorio Pierre Auger, Malarg\"ue, Argentina
\item[$^{11}$] Observatorio Pierre Auger and Comisi\'on Nacional de Energ\'\i{}a At\'omica, Malarg\"ue, Argentina
\item[$^{12}$] Universidad Tecnol\'ogica Nacional -- Facultad Regional Buenos Aires, Buenos Aires, Argentina
\item[$^{13}$] University of Adelaide, Adelaide, S.A., Australia
\item[$^{14}$] Universit\'e Libre de Bruxelles (ULB), Brussels, Belgium
\item[$^{15}$] Vrije Universiteit Brussels, Brussels, Belgium
\item[$^{16}$] Centro Brasileiro de Pesquisas Fisicas, Rio de Janeiro, RJ, Brazil
\item[$^{17}$] Centro Federal de Educa\c{c}\~ao Tecnol\'ogica Celso Suckow da Fonseca, Petropolis, Brazil
\item[$^{18}$] Instituto Federal de Educa\c{c}\~ao, Ci\^encia e Tecnologia do Rio de Janeiro (IFRJ), Brazil
\item[$^{19}$] Universidade de S\~ao Paulo, Escola de Engenharia de Lorena, Lorena, SP, Brazil
\item[$^{20}$] Universidade de S\~ao Paulo, Instituto de F\'\i{}sica de S\~ao Carlos, S\~ao Carlos, SP, Brazil
\item[$^{21}$] Universidade de S\~ao Paulo, Instituto de F\'\i{}sica, S\~ao Paulo, SP, Brazil
\item[$^{22}$] Universidade Estadual de Campinas (UNICAMP), IFGW, Campinas, SP, Brazil
\item[$^{23}$] Universidade Estadual de Feira de Santana, Feira de Santana, Brazil
\item[$^{24}$] Universidade Federal de Campina Grande, Centro de Ciencias e Tecnologia, Campina Grande, Brazil
\item[$^{25}$] Universidade Federal do ABC, Santo Andr\'e, SP, Brazil
\item[$^{26}$] Universidade Federal do Paran\'a, Setor Palotina, Palotina, Brazil
\item[$^{27}$] Universidade Federal do Rio de Janeiro, Instituto de F\'\i{}sica, Rio de Janeiro, RJ, Brazil
\item[$^{28}$] Universidad de Medell\'\i{}n, Medell\'\i{}n, Colombia
\item[$^{29}$] Universidad Industrial de Santander, Bucaramanga, Colombia
\item[$^{30}$] Charles University, Faculty of Mathematics and Physics, Institute of Particle and Nuclear Physics, Prague, Czech Republic
\item[$^{31}$] Institute of Physics of the Czech Academy of Sciences, Prague, Czech Republic
\item[$^{32}$] Palacky University, Olomouc, Czech Republic
\item[$^{33}$] CNRS/IN2P3, IJCLab, Universit\'e Paris-Saclay, Orsay, France
\item[$^{34}$] Laboratoire de Physique Nucl\'eaire et de Hautes Energies (LPNHE), Sorbonne Universit\'e, Universit\'e de Paris, CNRS-IN2P3, Paris, France
\item[$^{35}$] Univ.\ Grenoble Alpes, CNRS, Grenoble Institute of Engineering Univ.\ Grenoble Alpes, LPSC-IN2P3, 38000 Grenoble, France
\item[$^{36}$] Universit\'e Paris-Saclay, CNRS/IN2P3, IJCLab, Orsay, France
\item[$^{37}$] Bergische Universit\"at Wuppertal, Department of Physics, Wuppertal, Germany
\item[$^{38}$] Karlsruhe Institute of Technology (KIT), Institute for Experimental Particle Physics, Karlsruhe, Germany
\item[$^{39}$] Karlsruhe Institute of Technology (KIT), Institut f\"ur Prozessdatenverarbeitung und Elektronik, Karlsruhe, Germany
\item[$^{40}$] Karlsruhe Institute of Technology (KIT), Institute for Astroparticle Physics, Karlsruhe, Germany
\item[$^{41}$] RWTH Aachen University, III.\ Physikalisches Institut A, Aachen, Germany
\item[$^{42}$] Universit\"at Hamburg, II.\ Institut f\"ur Theoretische Physik, Hamburg, Germany
\item[$^{43}$] Universit\"at Siegen, Department Physik -- Experimentelle Teilchenphysik, Siegen, Germany
\item[$^{44}$] Gran Sasso Science Institute, L'Aquila, Italy
\item[$^{45}$] INFN Laboratori Nazionali del Gran Sasso, Assergi (L'Aquila), Italy
\item[$^{46}$] INFN, Sezione di Catania, Catania, Italy
\item[$^{47}$] INFN, Sezione di Lecce, Lecce, Italy
\item[$^{48}$] INFN, Sezione di Milano, Milano, Italy
\item[$^{49}$] INFN, Sezione di Napoli, Napoli, Italy
\item[$^{50}$] INFN, Sezione di Roma ``Tor Vergata'', Roma, Italy
\item[$^{51}$] INFN, Sezione di Torino, Torino, Italy
\item[$^{52}$] Istituto di Astrofisica Spaziale e Fisica Cosmica di Palermo (INAF), Palermo, Italy
\item[$^{53}$] Osservatorio Astrofisico di Torino (INAF), Torino, Italy
\item[$^{54}$] Politecnico di Milano, Dipartimento di Scienze e Tecnologie Aerospaziali , Milano, Italy
\item[$^{55}$] Universit\`a del Salento, Dipartimento di Matematica e Fisica ``E.\ De Giorgi'', Lecce, Italy
\item[$^{56}$] Universit\`a dell'Aquila, Dipartimento di Scienze Fisiche e Chimiche, L'Aquila, Italy
\item[$^{57}$] Universit\`a di Catania, Dipartimento di Fisica e Astronomia ``Ettore Majorana``, Catania, Italy
\item[$^{58}$] Universit\`a di Milano, Dipartimento di Fisica, Milano, Italy
\item[$^{59}$] Universit\`a di Napoli ``Federico II'', Dipartimento di Fisica ``Ettore Pancini'', Napoli, Italy
\item[$^{60}$] Universit\`a di Palermo, Dipartimento di Fisica e Chimica ''E.\ Segr\`e'', Palermo, Italy
\item[$^{61}$] Universit\`a di Roma ``Tor Vergata'', Dipartimento di Fisica, Roma, Italy
\item[$^{62}$] Universit\`a Torino, Dipartimento di Fisica, Torino, Italy
\item[$^{63}$] Benem\'erita Universidad Aut\'onoma de Puebla, Puebla, M\'exico
\item[$^{64}$] Unidad Profesional Interdisciplinaria en Ingenier\'\i{}a y Tecnolog\'\i{}as Avanzadas del Instituto Polit\'ecnico Nacional (UPIITA-IPN), M\'exico, D.F., M\'exico
\item[$^{65}$] Universidad Aut\'onoma de Chiapas, Tuxtla Guti\'errez, Chiapas, M\'exico
\item[$^{66}$] Universidad Michoacana de San Nicol\'as de Hidalgo, Morelia, Michoac\'an, M\'exico
\item[$^{67}$] Universidad Nacional Aut\'onoma de M\'exico, M\'exico, D.F., M\'exico
\item[$^{68}$] Institute of Nuclear Physics PAN, Krakow, Poland
\item[$^{69}$] University of \L{}\'od\'z, Faculty of High-Energy Astrophysics,\L{}\'od\'z, Poland
\item[$^{70}$] Laborat\'orio de Instrumenta\c{c}\~ao e F\'\i{}sica Experimental de Part\'\i{}culas -- LIP and Instituto Superior T\'ecnico -- IST, Universidade de Lisboa -- UL, Lisboa, Portugal
\item[$^{71}$] ``Horia Hulubei'' National Institute for Physics and Nuclear Engineering, Bucharest-Magurele, Romania
\item[$^{72}$] Institute of Space Science, Bucharest-Magurele, Romania
\item[$^{73}$] Center for Astrophysics and Cosmology (CAC), University of Nova Gorica, Nova Gorica, Slovenia
\item[$^{74}$] Experimental Particle Physics Department, J.\ Stefan Institute, Ljubljana, Slovenia
\item[$^{75}$] Universidad de Granada and C.A.F.P.E., Granada, Spain
\item[$^{76}$] Instituto Galego de F\'\i{}sica de Altas Enerx\'\i{}as (IGFAE), Universidade de Santiago de Compostela, Santiago de Compostela, Spain
\item[$^{77}$] IMAPP, Radboud University Nijmegen, Nijmegen, The Netherlands
\item[$^{78}$] Nationaal Instituut voor Kernfysica en Hoge Energie Fysica (NIKHEF), Science Park, Amsterdam, The Netherlands
\item[$^{79}$] Stichting Astronomisch Onderzoek in Nederland (ASTRON), Dwingeloo, The Netherlands
\item[$^{80}$] Universiteit van Amsterdam, Faculty of Science, Amsterdam, The Netherlands
\item[$^{81}$] Case Western Reserve University, Cleveland, OH, USA
\item[$^{82}$] Colorado School of Mines, Golden, CO, USA
\item[$^{83}$] Department of Physics and Astronomy, Lehman College, City University of New York, Bronx, NY, USA
\item[$^{84}$] Michigan Technological University, Houghton, MI, USA
\item[$^{85}$] New York University, New York, NY, USA
\item[$^{86}$] University of Chicago, Enrico Fermi Institute, Chicago, IL, USA
\item[$^{87}$] University of Delaware, Department of Physics and Astronomy, Bartol Research Institute, Newark, DE, USA
\item[] -----
\item[$^{a}$] Max-Planck-Institut f\"ur Radioastronomie, Bonn, Germany
\item[$^{b}$] also at Kapteyn Institute, University of Groningen, Groningen, The Netherlands
\item[$^{c}$] School of Physics and Astronomy, University of Leeds, Leeds, United Kingdom
\item[$^{d}$] Fermi National Accelerator Laboratory, Fermilab, Batavia, IL, USA
\item[$^{e}$] Pennsylvania State University, University Park, PA, USA
\item[$^{f}$] Colorado State University, Fort Collins, CO, USA
\item[$^{g}$] Louisiana State University, Baton Rouge, LA, USA
\item[$^{h}$] now at Graduate School of Science, Osaka Metropolitan University, Osaka, Japan
\item[$^{i}$] Institut universitaire de France (IUF), France
\item[$^{j}$] now at Technische Universit\"at Dortmund and Ruhr-Universit\"at Bochum, Dortmund and Bochum, Germany
\end{description}

\section*{Acknowledgments}

\begin{sloppypar}
The successful installation, commissioning, and operation of the Pierre
Auger Observatory would not have been possible without the strong
commitment and effort from the technical and administrative staff in
Malarg\"ue. We are very grateful to the following agencies and
organizations for financial support:
\end{sloppypar}

\begin{sloppypar}
Argentina -- Comisi\'on Nacional de Energ\'\i{}a At\'omica; Agencia Nacional de
Promoci\'on Cient\'\i{}fica y Tecnol\'ogica (ANPCyT); Consejo Nacional de
Investigaciones Cient\'\i{}ficas y T\'ecnicas (CONICET); Gobierno de la
Provincia de Mendoza; Municipalidad de Malarg\"ue; NDM Holdings and Valle
Las Le\~nas; in gratitude for their continuing cooperation over land
access; Australia -- the Australian Research Council; Belgium -- Fonds
de la Recherche Scientifique (FNRS); Research Foundation Flanders (FWO),
Marie Curie Action of the European Union Grant No.~101107047; Brazil --
Conselho Nacional de Desenvolvimento Cient\'\i{}fico e Tecnol\'ogico (CNPq);
Financiadora de Estudos e Projetos (FINEP); Funda\c{c}\~ao de Amparo \`a
Pesquisa do Estado de Rio de Janeiro (FAPERJ); S\~ao Paulo Research
Foundation (FAPESP) Grants No.~2019/10151-2, No.~2010/07359-6 and
No.~1999/05404-3; Minist\'erio da Ci\^encia, Tecnologia, Inova\c{c}\~oes e
Comunica\c{c}\~oes (MCTIC); Czech Republic -- GACR 24-13049S, CAS LQ100102401,
MEYS LM2023032, CZ.02.1.01/0.0/0.0/16{\textunderscore}013/0001402,
CZ.02.1.01/0.0/0.0/18{\textunderscore}046/0016010 and
CZ.02.1.01/0.0/0.0/17{\textunderscore}049/0008422 and CZ.02.01.01/00/22{\textunderscore}008/0004632;
France -- Centre de Calcul IN2P3/CNRS; Centre National de la Recherche
Scientifique (CNRS); Conseil R\'egional Ile-de-France; D\'epartement
Physique Nucl\'eaire et Corpusculaire (PNC-IN2P3/CNRS); D\'epartement
Sciences de l'Univers (SDU-INSU/CNRS); Institut Lagrange de Paris (ILP)
Grant No.~LABEX ANR-10-LABX-63 within the Investissements d'Avenir
Programme Grant No.~ANR-11-IDEX-0004-02; Germany -- Bundesministerium
f\"ur Bildung und Forschung (BMBF); Deutsche Forschungsgemeinschaft (DFG);
Finanzministerium Baden-W\"urttemberg; Helmholtz Alliance for
Astroparticle Physics (HAP); Helmholtz-Gemeinschaft Deutscher
Forschungszentren (HGF); Ministerium f\"ur Kultur und Wissenschaft des
Landes Nordrhein-Westfalen; Ministerium f\"ur Wissenschaft, Forschung und
Kunst des Landes Baden-W\"urttemberg; Italy -- Istituto Nazionale di
Fisica Nucleare (INFN); Istituto Nazionale di Astrofisica (INAF);
Ministero dell'Universit\`a e della Ricerca (MUR); CETEMPS Center of
Excellence; Ministero degli Affari Esteri (MAE), ICSC Centro Nazionale
di Ricerca in High Performance Computing, Big Data and Quantum
Computing, funded by European Union NextGenerationEU, reference code
CN{\textunderscore}00000013; M\'exico -- Consejo Nacional de Ciencia y Tecnolog\'\i{}a
(CONACYT) No.~167733; Universidad Nacional Aut\'onoma de M\'exico (UNAM);
PAPIIT DGAPA-UNAM; The Netherlands -- Ministry of Education, Culture and
Science; Netherlands Organisation for Scientific Research (NWO); Dutch
national e-infrastructure with the support of SURF Cooperative; Poland
-- Ministry of Education and Science, grants No.~DIR/WK/2018/11 and
2022/WK/12; National Science Centre, grants No.~2016/22/M/ST9/00198,
2016/23/B/ST9/01635, 2020/39/B/ST9/01398, and 2022/45/B/ST9/02163;
Portugal -- Portuguese national funds and FEDER funds within Programa
Operacional Factores de Competitividade through Funda\c{c}\~ao para a Ci\^encia
e a Tecnologia (COMPETE); Romania -- Ministry of Research, Innovation
and Digitization, CNCS-UEFISCDI, contract no.~30N/2023 under Romanian
National Core Program LAPLAS VII, grant no.~PN 23 21 01 02 and project
number PN-III-P1-1.1-TE-2021-0924/TE57/2022, within PNCDI III; Slovenia
-- Slovenian Research Agency, grants P1-0031, P1-0385, I0-0033, N1-0111;
Spain -- Ministerio de Ciencia e Innovaci\'on/Agencia Estatal de
Investigaci\'on (PID2019-105544GB-I00, PID2022-140510NB-I00 and
RYC2019-027017-I), Xunta de Galicia (CIGUS Network of Research Centers,
Consolidaci\'on 2021 GRC GI-2033, ED431C-2021/22 and ED431F-2022/15),
Junta de Andaluc\'\i{}a (SOMM17/6104/UGR and P18-FR-4314), and the European
Union (Marie Sklodowska-Curie 101065027 and ERDF); USA -- Department of
Energy, Contracts No.~DE-AC02-07CH11359, No.~DE-FR02-04ER41300,
No.~DE-FG02-99ER41107 and No.~DE-SC0011689; National Science Foundation,
Grant No.~0450696, and NSF-2013199; The Grainger Foundation; Marie
Curie-IRSES/EPLANET; European Particle Physics Latin American Network;
and UNESCO.
\end{sloppypar}

}

\end{document}